\documentclass[10pt,onecolumn,titlepage]{article}
\pdfoutput=1
\usepackage{scicite}

\usepackage{times}

\usepackage{color}
\usepackage{amsmath}
\usepackage{amssymb}
\usepackage[pdftex]{graphicx}
\usepackage{setspace}

\newenvironment{sciabstract}{%
\begin{quote} \bf}
{\end{quote}}

\newcounter{lastnote}

\title{The role of electron-electron interactions in two-dimensional \\ Dirac fermions}

\author
{Ho-Kin Tang,$^{1,2}$ J. N. Leaw,$^{1,2}$ J. N. B. Rodrigues,$^{1,2}$ I. F. Herbut,$^{3}$\\ P. Sengupta,$^{1,4}$ F. F. Assaad,$^{5}$ and S. Adam$^{1,2,6,\ast}$\\
\\
\normalsize{$^{1}$Centre for Advanced 2D Materials, National University of Singapore,}\\
\normalsize{6 Science Drive 2, 117546 Singapore.}\\
\normalsize{$^{2}$Department of Physics, Faculty of Science, National University of Singapore,}\\
\normalsize{2 Science Drive 3, 117542  Singapore.}\\
\normalsize{$^{3}$Department of Physics, Simon Fraser University,}\\
\normalsize{Burnaby, British Columbia V5A 1S6, Canada.}\\
\normalsize{$^{4}$School of Physical and Mathematical Sciences, Nanyang Technological University,}\\
\normalsize{21 Nanyang Link, 637371 Singapore.}\\
\normalsize{$^{5}$Institut f\"ur Theoretische Physik und Astrophysik, Universit\"at W\"urzburg,}\\
\normalsize{Am Hubland, D-97074 W\"urzburg, Germany.}\\
\normalsize{$^{6}$Yale-NUS College, 16 College Ave West, 138527 Singapore.}\\
\\
\normalsize{$^\ast$To whom correspondence should be addressed; E-mail: shaffique.adam@yale-nus.edu.sg}
}

\date{}

\begin{document} 

%\baselineskip24pt
%%%%%% Changed %%%%%%
\baselineskip16pt
%%%%%% Changed %%%%%%

\maketitle


\begin{sciabstract}
The role of electron-electron interactions on two-dimensional Dirac fermions remains enigmatic.  Using a combination of nonperturbative numerical and analytical techniques that incorporate both the contact and long-range parts of the Coulomb interaction, we identify the two previously discussed regimes: a Gross-Neveu transition to a strongly correlated Mott insulator, and a semi-metallic state with a logarithmically diverging Fermi velocity accurately described by the random phase approximation.  Most interestingly, experimental realizations of Dirac fermions span the crossover between these two regimes providing the physical mechanism that masks this velocity divergence.  We explain several long-standing mysteries including why the observed Fermi velocity in graphene is consistently about 20 percent larger than the best values calculated using {\it ab initio} and why graphene on different substrates show different behavior.
\end{sciabstract}

% ----------------------------------------------------------
%   Introduction
% ----------------------------------------------------------
%\section*{Introduction}
\subsection*{History of interacting Dirac fermions}
In 1952, Freeman Dyson made the argument that all theoretical methods for solving problems 
in quantum electrodynamics are in the form of a perturbative asymptotic series expansion in the fine-structure 
constant~$\alpha$, where the Coulomb potential is of the form $\alpha/r$ where $r$ is the distance between two electrons. Further, he showed that any such series has a fundamental divergence making such solutions uncontrolled beyond 
the order of perturbation theory given by the inverse fine-structure constant~\cite{Dyson_PR:1952}. Since for quantum electrodynamics, 
$\alpha \approx 1/137$, this divergence is rather academic since the perturbative expansion approximates the 
true result for roughly the first 137 terms.  For condensed matter realizations of two-dimensional Dirac 
fermions, the long-range Coulomb coupling constant $\alpha \sim 1$, which implies that any perturbative theory 
is potentially uncontrolled even at first order~\cite{Barnes_PRB:2014}.  This formal constraint has not precluded the 
development of perturbation theory for such systems.  This is particularly problematic because the first-order 
perturbation theory gives rather peculiar results.  To first order in $\alpha$, the inverse coupling constant 
(proportional to the Fermi velocity) itself diverges both in the infrared (long distances) and ultraviolet (small 
distances).  Introducing a lattice-scale fixes the ultraviolet divergence, but not the infrared one~\cite{Gonzalez_NPB:1994,Gonzalez_PRB:1999}.

This divergence is what lead Ye and Sachdev in 1998 to describe the effects of long-range electron-electron interactions 
as {\it ``dangerously irrelevant"}, by which they meant that although the system flows under the renormalization 
group to a non-interacting theory, physical observables are strongly renormalized by the Coulomb interaction 
\cite{Ye_PRL:1998}.  Since then, dozens of theoretical works have confirmed this basic picture that 
in the absence of disorder, and precisely at half-filling, the role of long-range Coulomb interactions is to 
renormalize the electron Fermi velocity to infinity (limited, only by the speed of light if one includes 
a dynamical interaction), and reducing the quasi-particle residue to zero, thereby signaling a non-Fermi liquid metallic state 
\cite{Kotov_RMP:2012}.  For the purposes of this work, none of the developments following Ye and Sachdev can be 
distinguished by our numerical results, and they make only small differences for our predictions for experiments 
\cite{SuppInfo}.

In a parallel development, the Hubbard model on a honeycomb lattice provided a realization at low energy of Dirac 
fermions interacting through a short-range contact interaction $U$.  Increasing the 
short-range interaction results in a quantum phase transition at $U=U_c$ from a semi-metal to an anti-ferromagnetic 
Mott insulator. In 2006, one of us predicted that this phase transition is of the Gross-Neveu universality class \cite{Herbut_PRL:2006}, which was recently confirmed numerically~\cite{Toldin_PRB:2015,Otsuka_PRX:2016}.  In diagrammatic perturbation theory, typically both the Fermi velocity and quasi-particle residue vanish 
at a quantum phase transition.  However, for the Gross-Neveu critical point, the Fermi velocity 
remains finite (despite the vanishing of the quasiparticle residue), and this suppression of the Fermi velocity to a finite 
value has been observed numerically~\cite{Sorella_private}. For $U \ll U_c$, renormalization group studies have shown that the Fermi velocity remains unchanged due to interactions~\cite{Giuliani_CMP:2009}, and in our numerics below, we verify all these features for the Hubbard model on a honeycomb lattice.  However, since two-dimensional Dirac fermions are unable to screen the long-range Coulomb potential,  it is widely believed \cite{DasSarma_RMP:2011,Kotov_RMP:2012} that this onsite Hubbard model would have limited applicability to experiments done in real materials.

Experimentally, two-dimensional Dirac fermions can be realized in a variety of condensed matter systems including on 
the surfaces of 3D topological insulators \cite{Hsieh_Nature:2008,Kim_NatPhys:2012} and in artificial graphene made from quantum corrals of carbon monoxide arranged in honeycomb lattice on a copper substrate~\cite{Gomes_NAT:2012}.  Other examples are discussed in the supplemental material~\cite{SuppInfo}.  For concreteness, we focus our attention on graphene, the most 
studied and versatile realization of 2D Dirac fermions. While experiments have been unable to realize the precise configuration 
necessary to probe this strange interacting metallic state -- that features, simultaneously having the electron quasi-particles 
moving at the speed of light, while their quasiparticle character smears away -- several studies in ultra-clean graphene 
including the role of magnetic field on the transport \cite{Elias_NatPhys:2011}, infrared spectroscopy \cite{Li_NatPhys:2008}, 
capacitance \cite{Yu_PNAS:2013}, ARPES \cite{Hwang_SciRep:2012}, tunneling spectra \cite{Chae_PRL:2012} and Raman scattering 
\cite{Faugeras_PRL:2015} all reveal a clear breakdown of the non-interacting theory.

In this work, we address the competing effects of short-range and long-range parts of any realistic model of the Coulomb interaction.  We use a non-perturbative, numerically exact, projective quantum Monte Carlo (QMC) method to study the evolution of physical observables in a controllable manner. We find that in the regime dominated by long-range interactions, there is an enhancement 
of Fermi velocity consistent with perturbation theory.  Conversely, close to the phase transition dominated by short-range 
interactions, we find a suppression of Fermi velocity and a collapse of the numerical data for different values of the ratio 
between long-range and short-range interactions. Our numerical results interpolate between these two limits and are valid for 
all interaction strengths, but are constrained only by the finite system sizes in our simulations.  Therefore, we use a renormalization 
group scheme to extrapolate the quantum Monte Carlo results to experimentally relevant energy-scales where we predict that 
observables will depend on both the short-range and long-range components of the Coulomb interaction as well as the energy-scale of the observation (and as we explain below, all these can be tuned in current experiments).  Moreover, we find that the lattice-scale not only regularizes the ultraviolet divergence of the Fermi velocity, but also can make the infrared divergence unnoticeable in the experimental window.

%%%%%%%%%%%%%%%%%%%%%%%%%%%%%%%%%%%
%%%%%%%  The phase diagram  %%%%%%%
%%%%%%%%%%%%%%%%%%%%%%%%%%%%%%%%%%%
\begin{figure}[htp!]
  \centering
  \includegraphics[width=\columnwidth]{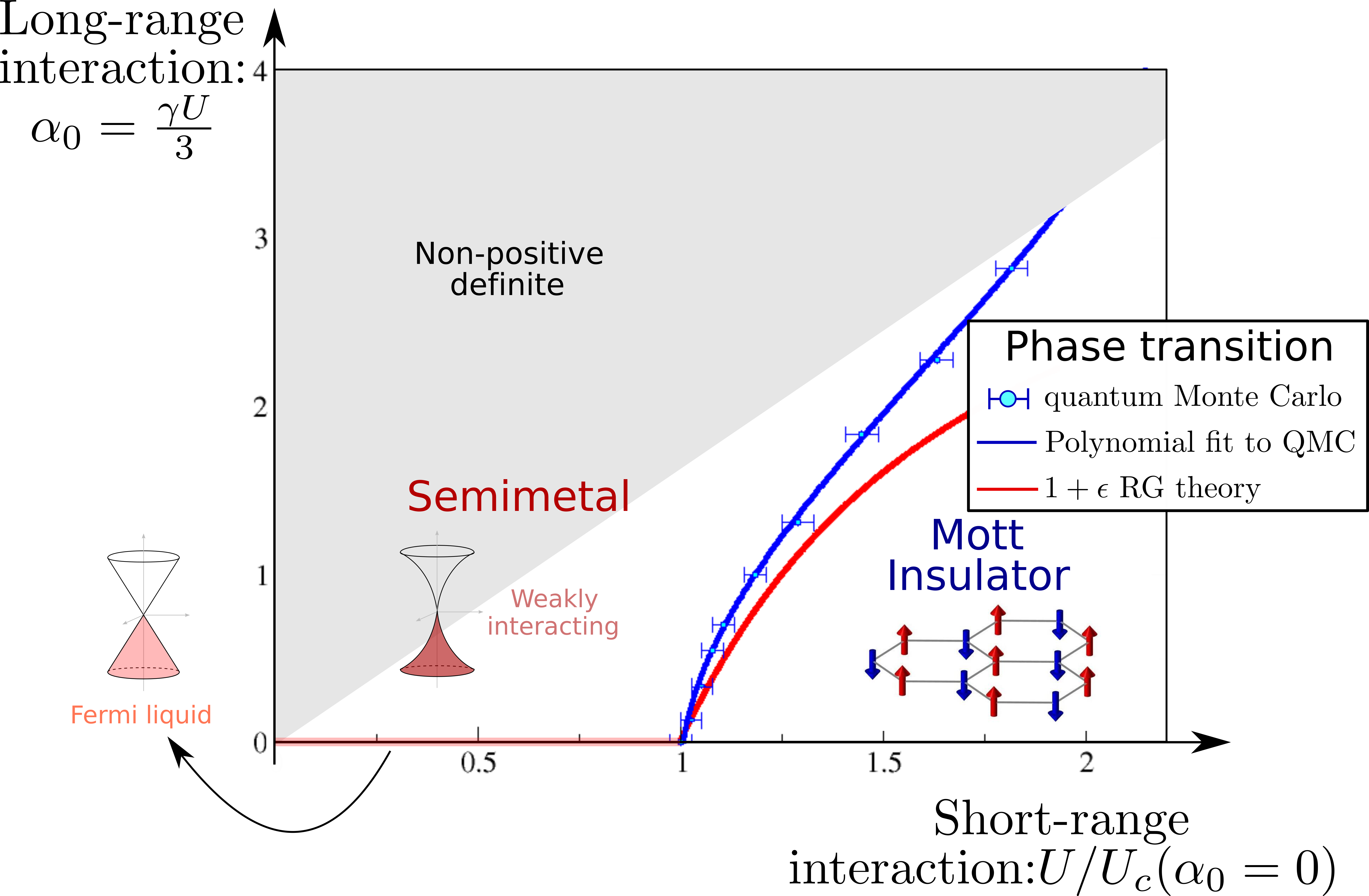}
  \caption{\textbf{Phase diagram for fermions on the honeycomb lattice with competing short-range and long-range Coulomb interactions.} For any value of long-range interaction $\alpha_0$, there is a critical value of the short-range interaction $U_c(\alpha_0)$ calculated using quantum Monte Carlo (data points), for which the system undergoes a quantum phase transition to the Mott insulator.  In the presence of long-range interactions, a larger value of onsite interactions is required to reach the quantum phase transition.  The phase diagram can be understood by solving the renormalization group flow equations (red curve) including both onsite and nearest neighbour interactions, where the effective onsite interactions are reduced by the long-range Coulomb tail.  Solid blue line is a quartic interpolation.  The shaded window shows the region inaccessible to our numerical method~\protect{\cite{SuppInfo}}.}
  \label{fig:Fig1_PhaseDiagram}
\end{figure}

% ----------------------------------------------------------
%   Model and methods
% ----------------------------------------------------------

\subsection*{Theoretical model}
In order to accomplish this, we study interacting fermions on a honeycomb lattice with competing short- and long-range 
interactions. The model is described by the Hamiltonian 
$\hat{H} = - t \sum_{\langle i j \rangle, \sigma} \big( \hat{c}_{i\sigma}^{\dagger} \hat{c}_{j\sigma} + h.c. \big )
+ \frac{1}{2} \sum_{i, j} (\hat{n}_{i}-1) \mathcal{V}_{i j} (\hat{n}_{j}-1) \,$
where $\hat{c}_{i \sigma}^{\dagger}$ ($\hat{c}_{i \sigma}$) creates (annihilates) an electron of spin $\sigma = \uparrow 
\downarrow$ at position $\mathbf{r}_{i}$ and $\hat{n}_{i} = \sum_\sigma\hat{c}_{i \sigma}^{\dagger} \hat{c}_{i \sigma}$ gives the electron density at position $\mathbf{r}_{i}$. The interaction between electrons
is parameterized as $\mathcal{V}_{ij}$ where the onsite interaction for electrons with different spins 
$\mathcal{V}_{ii}=U$, while the long-range interaction depends on the distance between the electrons 
$r_{ij}=2 a \left|\mathbf{r}_i-\mathbf{r}_j\right|/3$ as $\mathcal{V}_{ij}=\alpha_0/r_{ij}$, where $a$ is the lattice constant.  Using our quantum Monte Carlo method, we are able to map out the phase diagram (Fig.~\ref{fig:Fig1_PhaseDiagram}) of this model showing the competing effects of the short-range and long-range parts of the Coulomb potential.  The inclusion of the long-range component was made possible by recent developments in lattice quantum chromodynamics~\cite{Brower_POS:2012}, which we adapt \cite{Tang_PRL:2015} for our purposes here.  Contrary to some expectations in the literature~\cite{DasSarma_RMP:2011,Kotov_RMP:2012}, we find that the critical onsite interaction $U_c(\alpha_0)$ increases when the long-range Coulomb interaction is included.  This can be understood in two ways.  First, as one increases the long-range Coulomb tail, the effective onsite potential decreases (one can think of this, qualitatively, as the difference between the onsite potential and the nearest neighbour potential). Therefore, with inclusion of the long-range piece, one needs a larger onsite potential to get the same effective critical Hubbard potential\cite{Schuler_PRL:2013}.  Second, while the Hubbard potential favours an antiferromagnetic ground state, the nearest neighbour potential favours instead a charge density wave ground state.  By including the long-range piece, one needs a larger onsite potential to stabilize the antiferromagnetic phase.  Both these pictures are confirmed in our renormalization group calculation (shown as the red curve in the figure) and discussed in the supplemental material~\cite{SuppInfo}. The shaded region in the figure is inaccessible in our numerics because the quantum Monte Carlo does not converge here due to a breakdown of the method we use to incorporate the long-range Coulomb potential. In light of this, the fact that the phase transition line goes to the right when $\alpha_0$ is increased is fortuitous for us, since this opens up a large parameter space for our simulations.

%%%%%%%%%%%%%%%%%%%%%%%%%%%%%%%%%%%%%%%%%%
%%%%%%% Fermi velocity main figure %%%%%%%
%%%%%%%%%%%%%%%%%%%%%%%%%%%%%%%%%%%%%%%%%%
\begin{figure}[htp]
  \centering
  \includegraphics[width=\columnwidth]{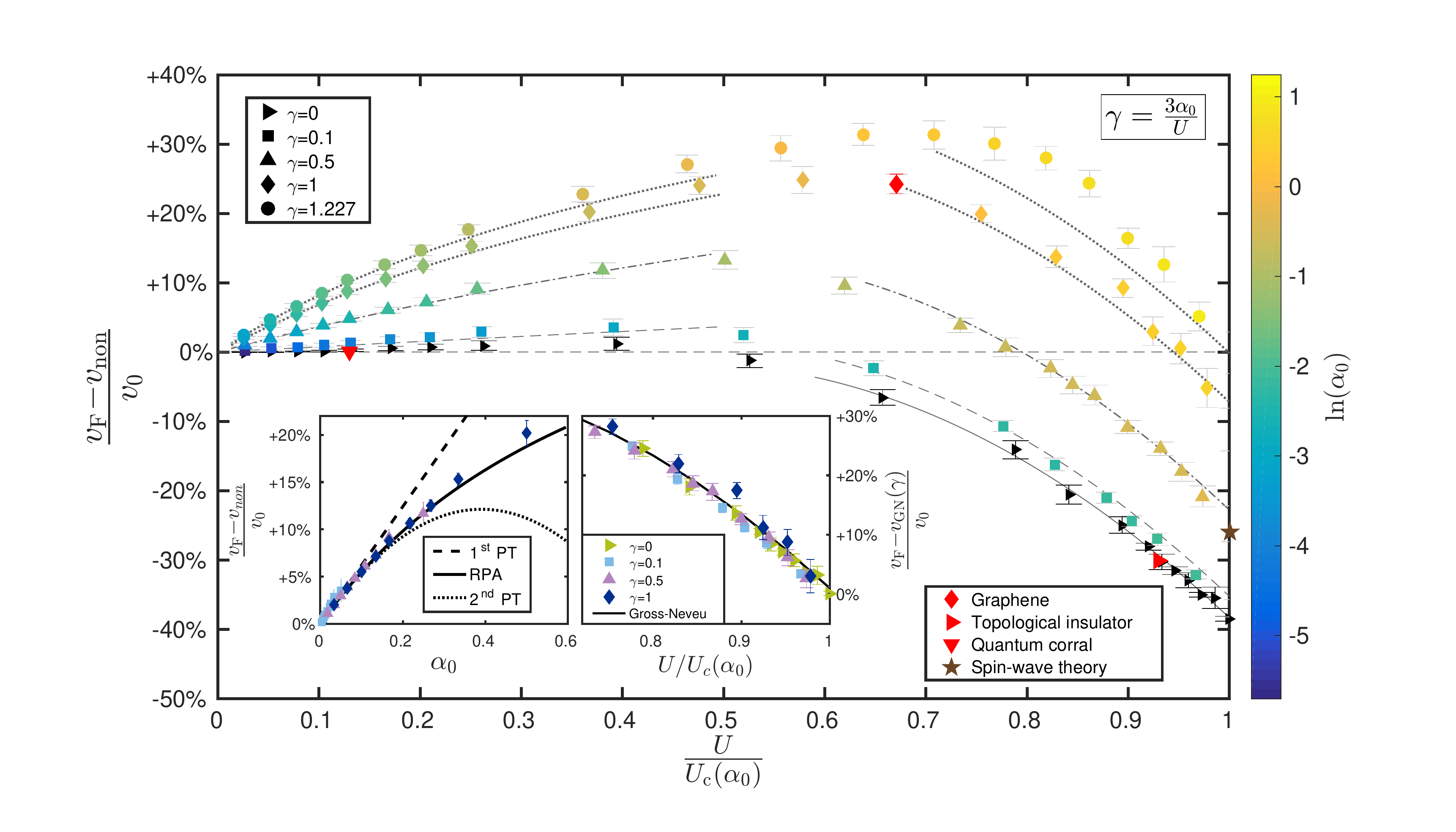}
  \caption{\textbf{Dirac fermion Fermi velocity renormalized by electron-electron interactions.}  Projective quantum Monte Carlo results for different short-range ($U$) and long-range ($\alpha_0$) components of the Coulomb interaction.  Small $U/U_c(\alpha_0)$ defines the weak coupling regime, where Monte Carlo data for different ratios $\gamma$ of the long-range and short-range components collapse as a function of $\alpha_0$.  Here electron-electron interactions {\it enhance} the Fermi velocity in agreement with the random phase approximation (left inset). A metal-to-Mott insulator phase transition of the Gross-Neveu universality class occurs at $U = U_c(\alpha_0)$, where a {\it suppression} of Fermi velocity can be understood as the coupling between Dirac fermions and the bosonic excitations of the nascent antiferromagnetic state (the brown star is an estimate of this Fermi velocity suppression determined using spin-wave theory).  The right inset shows that QMC data for the change in Fermi velocity from the value at the Gross-Neveu point collapses as one moves away from the phase transition.  Our numerics span the full cross-over between the weak coupling fixed point and Gross-Neveu critical point.  Estimates place topological insulators close to the phase transition, while quantum corral-like honeycomb lattices are in the weak coupling limit.  Graphene Dirac fermions lie somewhere in between these two regimes.}
  \label{fig:Fig3_vF-Main}
\end{figure}

Figure \ref{fig:Fig3_vF-Main} shows our main numerical results on the Fermi velocity, which is the defining property of the massless Dirac spectrum. We plot the renormalization of Fermi velocity $(v_{F}-v_{non})/v_{0}$ against the distance to the phase transition $U/U_{c}(\alpha_0)$. Here, $v_{F}$ is the interaction-renormalized Fermi velocity obtained in quantum Monte Carlo at the simulation scale $\Lambda_s \equiv ka=0.48$, $v_{non}$ is the tight-binding Fermi velocity at the simulation scale, and $v_{0}$ is the tight-binding Fermi velocity at the Dirac point.  Each data point in the figure is an extrapolation from four lattice sizes, each with an average of a hundred thousand quantum Monte Carlo sweeps.  The interacting Fermi velocity at the simulation scale, is obtained by first determining the convergent ground state value of the unequal time Green function $G_k(\tau)$ for large $\tau$, where the single exponential decay time, $\log\left[G_k(\tau)\right] \sim (E^{N+1}_{k,G.S.}-E^{N}_{0,G.S.}) \tau$ determines the first excitation energy of the system (defined here as the energy difference between the ground state of N+1 fermions with a total momentum $k$ and the ground state of N fermions with zero total momentum).  Full details of our quantum Monte Carlo scheme, data sets and the analysis are provided in the supplementary material~\cite{SuppInfo}.

Our numerical data shows that the Fermi velocity renormalization has strikingly different behaviour for $U/U_{c}(\alpha_0)\ll 1$ (which we call the `weak-coupling regime') and $U/U_{c}(\alpha_0) \lesssim 1$ which is in the vicinity of the Gross-Neveu critical point.  In the weak coupling regime, we observe an increase in the Fermi velocity and all the quantum Monte Carlo data for different ratios of short-range and long-range components collapse when plotted as a function of the long-range interaction $\alpha_0$ (see left inset).  In contrast, in the vicinity of the Gross-Neveu critical point, the Hubbard model ($\alpha_0 = 0$) shows a 40 percent decrease in Fermi velocity.  Even after including the long-range component of the Coulomb interaction, after subtracting the intercept of the Fermi velocity at the Gross Neveu critical point, all the numerical data collapse to the Hubbard model function form (see right inset).  This shows that interacting fermions on honeycomb lattice are governed by two very different fixed points, one controlled by the long-range interaction giving an {\it enhancement} in Fermi velocity, and the other governed by the short-range interaction giving a {\it suppression} of the Fermi velocity. An estimate for the realistic Coulomb potential in graphene places it in the crossover between these two regimes (see red diamond in the figure), while topological insulator $\mathrm{Bi}_2\mathrm{Se}_3$ is close  to the Mott transition and artificial graphene using quantum corrals is in the weak coupling regime~\cite{SuppInfo}.  Our numerical results span the full crossover between these two regimes.

%%%%%%%%%%%%%%%%%%%%%%%%%%%%%%%%%%%%%%%%%%
%%%%%%% Flow of the Fermi velocity %%%%%%%
%%%%%%%%%%%%%%%%%%%%%%%%%%%%%%%%%%%%%%%%%%
\begin{figure}[htp]
  \centering
  \includegraphics[width=\columnwidth]{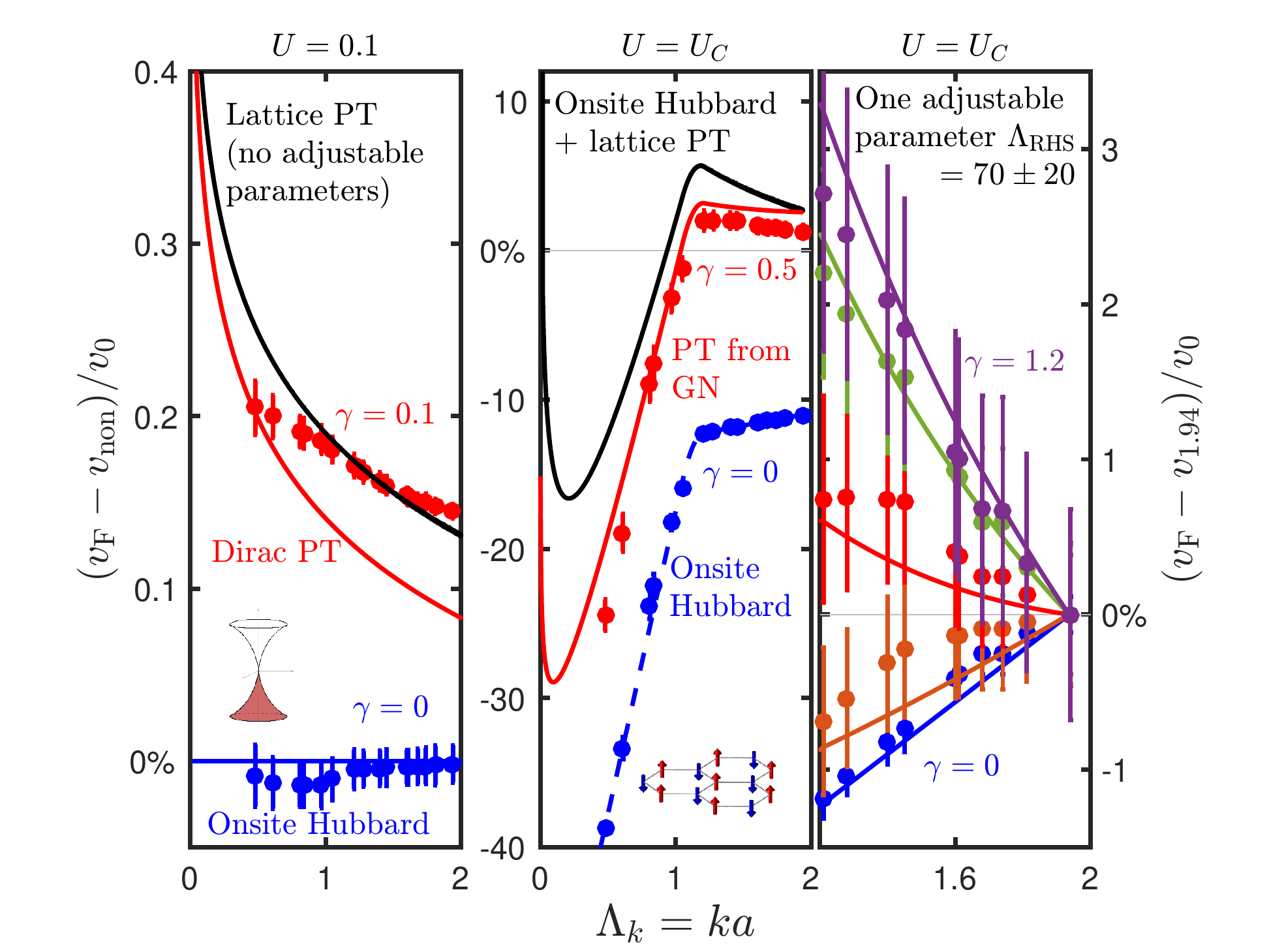}
  \caption{\textbf{Determining the renormalization group flow parameters from our quantum Monte Carlo data.}  Our simulations also provide data for momenta larger than $\Lambda_s$.  We exploit this scale dependence to determine scales smaller than what we can simulate.  Left panel shows representative data close to the weak coupling fixed point where the onsite Hubbard model (blue data) shows no observable change in Fermi velocity.  With long-range interactions, the Fermi velocity increases with decreasing momenta (red data) understood either using a continuum perturbation theory (red curve) or lattice perturbation theory with no adjustable parameter (black curve) which diverge logarithmically.  The Gross-Neveu critical point in the middle panel is very different.  Here, neither the Hubbard model (blue data) nor the data including the long-range Coulomb interaction (red data) show a logarithmic divergence at small $\Lambda_k$.  A phenomenological fit captures the onsite Hubbard model (blue dashed line).  The weak increase in renormalized Fermi velocity as one goes from $\Lambda_k = 2$ to $\Lambda_k =1$ is seen in both the black curve (lattice perturbation theory) and the QMC data with finite long-range interactions.  The red curve is obtained by fitting for this increase using a first order perturbation theory about the Gross-Neveu critical point (right panel).}
  \label{fig:Fig4_vF-flow}
\end{figure}

%%%%%%%%%%%%%%%%%%%%%%%%%%%%%%%%%%%%%%%%%%%%%
%%%%%%% Flow of the coupling constant %%%%%%%
%%%%%%%%%%%%%%%%%%%%%%%%%%%%%%%%%%%%%%%%%%%%%
\begin{figure}[htp]
  \centering
  \includegraphics[width=\columnwidth]{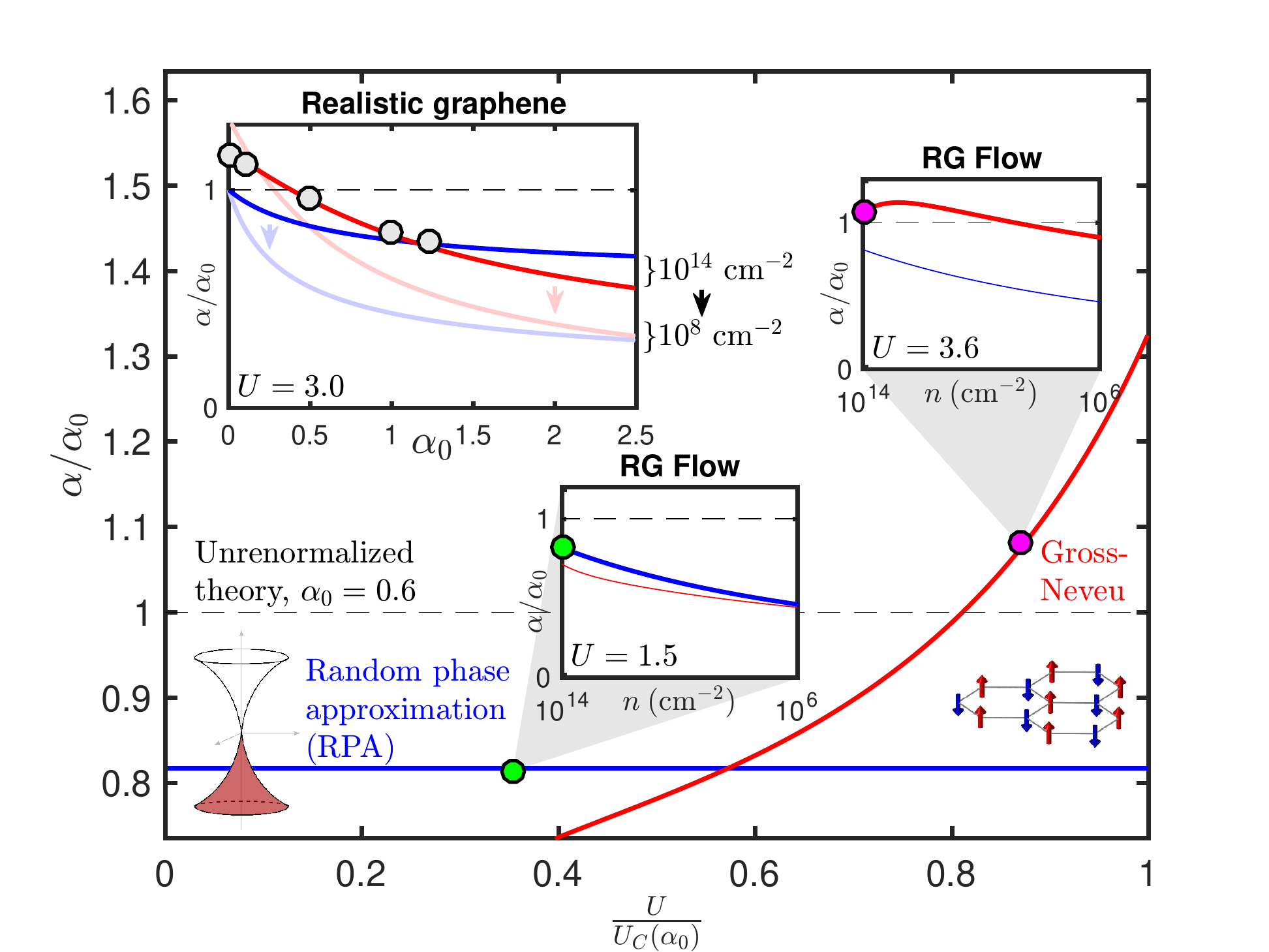}
  \caption{
  \textbf{Theoretical predictions for coupling constant renormalization for different experimental realizations of graphene.}  Main panel shows the simulation scale using the non-interacting value $\alpha_0 = 0.6$.  The blue curve shows the RPA suppression in coupling constant that has no dependence on the short-range interaction.  The red curve shows our Gross-Neveu phenomenological theory.  The green data taken from the weak coupling regime agrees with the RPA, and the associated inset shows that for experimentally realizable energy scales, the RG flows are similar for the two theories.  The magenta data is taken close to the Gross-Neveu critical point, where unlike the RPA, the theory has almost no renormalization in the experimental window.  This implies counter-intuitively that the lattice-scale component of the Coulomb potential also regularizes the long-distance divergence of the Fermi velocity.  The top inset uses parameters for realistic graphene.  For the most common realization of graphene on a dielectric substrate ($\alpha_0 \approx 0.6$), we offer new predictions of a weak suppression of coupling constant that slightly changes under renormalization.  Most surprising, we predict that for small $\alpha_0$ (e.g. graphene on metal substrates or for topological insulators) there should be an interaction enhancement of coupling constant that further increases upon renormalization.}
  \label{fig:Fig5_alpha-flow}
\end{figure}

The emergence of a stable ``weak-coupling'' fixed point and an unstable Gross-Neveu fixed point is anticipated by renormalization group studies.  In the weak-coupling regime, all the quantum Monte Carlo data (broken lines in the left-hand side of the main panel) can be reproduced using a one-parameter random phase approximation~(RPA) theory $v^\mathrm{RPA}\left(k\right) = v_{0}\left\{ 1+ \left[F_{1}\left(\alpha_0\right)-F_{0}\left(\alpha_0\right)\right]\ln\left(\frac{\Lambda}{\left|\Lambda_s\right|}\right)\right\}$, where $\Lambda_s$ is our numerical scale and $\Lambda = 6.2\pm 0.2$ is obtained from fitting the data for small $\alpha$.  The RPA functions $F_1(\alpha_0)$ and $F_0(\alpha_0)$ have been rederived several times in the literature (and are reproduced in the supplementary material~\cite{SuppInfo}).  Our non-perturbative  numerics, for the first time, verify this functional dependence on $\alpha_0$.  

While currently there is no analytical theory for the dependence on $U/U_c$ close to the Gross-Neveu fixed point (even for the pure Hubbard model without any long-range interactions), we can adequately describe the quantum Monte Carlo data using a phenomenological model that has 3 parameters for the Hubbard model data and one additional parameter for linear dependence on $\gamma = 3 \alpha_0/U$ of the Fermi velocity at the Gross-Neveu critical point $v_{GN}(\gamma)$. Defining $\epsilon = 1 - U/U_c(\alpha_0)$, we find that $(v_F - v_{non})/v_0 = C_0 + C_1 \epsilon + C_2 \epsilon^2 + m_{\gamma} \gamma$ with $C_0 = -0.384 \pm 0.002, C_1 = 1.35 \pm 0.05$, $C_2 = -1.2 \pm 0.1$ determined just from the $\alpha_0 = 0$ pure Hubbard model case, and the single additional parameter $m_\gamma =0.333 \pm 0.005$ captures the effects of the long-range Coulomb potential.  We can understand the suppression of the Fermi velocity at the Gross-Neveu fixed point in the following picture.  As the antiferromagnetic phase starts to emerge, there are bosonic Goldstone modes of this broken symmetry.  These slow modes couple to the Dirac fermions reducing their Fermi velocity.  We estimate this effect by calculating the bosonic velocity using large spin and large $U$ spin-wave theory, and then invoke a renormalization group argument~\cite{Roy_JHEP:2016} showing that at the critical point the bosonic and fermionic velocity converge in the infrared limit.  Although this analysis is approximate, it nonetheless does a reasonable job at estimating the reduced Fermi velocity as shown by the brown star in Fig.~\ref{fig:Fig3_vF-Main}.     

\subsection*{Flow beyond numerical scales}
The analysis so far has been at the simulation scale $\Lambda_s = 0.48$ that is limited by the largest system size we can simulate.  However, the experimental scale is set mostly by the degree of disorder in the system (currently scales as small as $\Lambda_k = 10^{-3}$ can be measured).  How then do we extrapolate our numerical findings to the experimental regime?  While we cannot numerically probe scales smaller than $\Lambda_s$, we can probe larger energy scales, thereby providing the inputs for a renormalization group flow from the numerical scale to the experimental scale.  This procedure is illustrated in Fig.~\ref{fig:Fig4_vF-flow}.  The left panel is for $U = 0.1$ (in the weakly coupling regime).  For the pure Hubbard model ($\alpha_0 = 0$), there is no renormalization of the Fermi velocity even at larger energy scales (blue circles).  For small $\alpha_0$ (shown in the figure is $\gamma = 3 \alpha_0/U = 0.1$ as a representative example) the numerical data decreases weakly with increasing $\Lambda_k$.  To show that this decrease is consistent with a logarithm we show two theoretical analysis.  The red curve shows the first order perturbation theory of a Dirac spectrum with the same single global parameter discussed already in Fig~\ref{fig:Fig3_vF-Main}.  We attribute the disagreement with the QMC to the fact that at such large energy scales the lattice model has significant non-linear terms.  To check this, we also solve numerically the first-order perturbation theory in $\alpha_0$ on a finite lattice.  Since the lattice is specified, there is no adjustable parameter in this calculation.  This is shown as the black curve in the figure and it agrees with the quantum Monte Carlo results at large energy scales.  In this method, we can go to lattice sizes as large as $1500 \times 1500$, and in the small $\Lambda_k$ window, the lattice perturbation theory results look similar to the logarithmic divergence in the continuum perturbation theory.  

The middle panel shows representative examples at the Gross-Neveu critical point.  The Hubbard model (blue circles) show that the renormalized Fermi velocity increases with energy scale.  The dashed line shows a phenomenological fitting function to capture these numerical results.  Including the long-range component, the dominant effect is an overall upward shift in the curve (as already discussed in Fig.~\ref{fig:Fig3_vF-Main}). However, at large $\Lambda_k$, we notice another subtle difference.  While the Hubbard model always shows the renormalized velocity monotonically increasing function of $\Lambda_k$, with long-range interactions the dependence is non-monotonic.  This observation suggests adding the parameter-free lattice perturbation theory to the Hubbard model with a single shift parameter obtained by fitting the full dataset of various $\gamma$ at the Gross-Neveu fixed point.  This gives the black curve in the middle panel.  The key insight is that while the effect of the logarithm is not observable at $\Lambda_s$, it can be extracted at \textit{larger} values of $\Lambda_k$.  We observe that the divergence in the QMC data is weaker than in the black curve.  Both this observation, and looking at the structure of the terms in a perturbation theory about the Gross-Neveu fixed point suggest fitting for the logarithm at large $\Lambda_k$~\cite{SuppInfo}.  The right panel shows a blow-up of the region around $\Lambda_k = 2$, where a single adjustable parameter $\Lambda_{RHS} = 70 \pm 20$ describes this increase in Fermi velocity at Gross-Neveu critical point for all our long-range data.  The red curve in the middle panel shows for $\gamma = 0.5$ the analytical first-order perturbation theory result (similarly shifted like the black curve), but including this logarithmic term to capture the decrease of velocity at large $\Lambda_k$. In addition, we observe a linear shift in $C_1(\Lambda_k) = C_1(0.48) - (0.92 \pm 0.06) (\Lambda_k - 0.48)$ where the slope was determined also by looking at the full QMC data set~\cite{SuppInfo}.  Therefore, combining this with the results of Fig~\ref{fig:Fig3_vF-Main} and Fig~\ref{fig:Fig4_vF-flow}, we can now extrapolate our QMC findings to any value of $\alpha_0$, $U$ and $\Lambda_k$, thereby making predictions for realistic Dirac fermions at any experimental scale.

\subsection*{Implications for experiments}
Our results on the role of electron-electron interactions apply to any Dirac fermion system for which one can define the short-range component of the Coulomb interaction $U$, its long-range tail $\alpha_0$ and the experimental probe energy scale $\Lambda_k$. However, in Fig.~\ref{fig:Fig5_alpha-flow}, we use our results to make predictions for graphene because the coupling constant for graphene Dirac fermions $\alpha_0 = e^2/(\kappa \hbar v_0)$ can be tuned using substrates of different dielectric constants $\kappa$.  The onsite potential is set to $U = 3.0$ following estimates made for graphene (see supplemental material~\cite{SuppInfo} for discussion of parameters for different condensed matter realizations of Dirac fermions). We use density units in the insets (instead of $\Lambda_k$) for ease of comparison with experimental results, although our numerical results only apply at half-filing.  For the most typical situation of graphene on a substrate, typical measured Fermi velocities are (1.1 \textendash~1.3) $\times 10^{6}~\mathrm{m/s}$~\cite{DasSarma_RMP:2011}, however, \textit{ab initio} calculations predict $0.87 \times 10^{6}~\mathrm{m/s}$ (see e.g. Ref.~\cite{Trevisanutto_PRL:2008}).  This significant discrepancy between theory and experiment has been largely unresolved in the literature -- a notable exception is Ref.~\cite{Trevisanutto_PRL:2008} who used an \textit{ab initio} DFT-GW calculation that agrees precisely with our result at our simulation scale, but their spectrum became unphysical for $n \lesssim 3 \times 10^{11}~\mathrm{cm}^{-2}$).  However, this interaction enhancement of the Fermi velocity (or equivalently suppression in coupling constant) can be seen directly in the top left inset of Fig.~\ref{fig:Fig5_alpha-flow}, where, for example, at $\alpha_0 = 1$, and typical experimental energy scale $n = 1 \times 10^{10}~\mathrm{cm}^{-2}$, we predict $\alpha = 0.65 \pm 0.03$ corresponding to Fermi velocity of $(1.34 \pm 0.07) \times 10^{6}~\mathrm{m/s}$.  

We predict that for topological insulator Bi$_2$Se$_3$ ($\alpha_0 \approx 0.05$) or for graphene on metallic substrates, 
interactions \textit{enhance} the coupling constant in the experimental window rather than the expected suppression.  This enhancement originates from the suppression of Fermi velocities in the Hubbard model at the Gross-Neveu fixed point.  The phenomenological theory predicts that the coupling constant is almost unrenormalized (e.g. for $\alpha_0 = 0.05$ the coupling changes by less than 5 percent over the entire experimental window \textendash the two small insets in Fig.~\ref{fig:Fig5_alpha-flow}).  Most surprising for sufficiently small $\alpha_0$, we predict that flowing $\Lambda_k$ closer to the Dirac point increases $\alpha$ rather than the expected decrease.  These predictions challenge the conventional wisdom on the role of electron-electron interactions in two-dimensional Dirac fermions.  On the other hand, for suspended graphene, or quantum corral-like graphene, we expect the experiments to be close to the weak-coupling fixed point and therefore, the random phase approximation should work well in this regime.   All our predictions can readily be tested with current experimental capabilities.

To conclude, we have used a non-perturbative numerically exact projective quantum Monte Carlo method that incorporates both the long-range and short-range components of the electron-electron interactions between Dirac fermions.  At our numerical energy scale, we clearly identify the weak coupling fixed point that is well described by the random phase approximation, and a metal-to-insulator transition characterized by Gross-Neveu exponents.  In the crossover between these two fixed points, the lattice scale of the numerics cures both the infrared divergence and the ultraviolet divergence predicted by perturbation theory for Dirac fermions.  Although no analytical theory exists for the Fermi velocity in the vicinity of the quantum critical point, we develop a phenomenological model that adequately captures our numerical data.  Using the energy scale dependence in our numerics, we set the parameters for a renormalization group procedure that allows us to extrapolate our results from the numerical scale to the experimental window.  In the experimental window we solve several long-standing mysteries including discrepancy between experiments and {\it ab initio} calculations, and why graphene on different substrates show different interaction-driven behaviour.  We predict that for sufficiently small coupling constant (such as in topological insulators or graphene on metallic substrates) the Fermi velocity should be suppressed rather than enhanced by electron-electron interactions.  Our work provides the first comprehensive treatment for the role of realistic Coulomb interaction between Dirac fermions and the relevance for their different experimental realizations.

\bibliographystyle{Science}
\bibliography{Graphene_e-e_Interactions_QMC}

\noindent We acknowledge allocation of computational resources at the CA2DM (Singapore) and the Gauss Centre for Supercomputing (SuperMUC at the  Leibniz Supercomputing Center).  This project was supported by the Singapore National Research Foundation (NRF-NRFF2012-01),  Deutsche Forschungsgemeinschaft (SFB 1170 ToCoTronics, project C01), NSERC of Canada,  and Singapore Ministry of Education (MOE2014-T2-1-112 and MOE2017-T2-1-130).  The full collection of data developed in this work is available at
\href{https://figshare.com/articles/Source\_data\_of\_Green\_s\_function\_and\_figure\_files\_/5131840/2}{
https://figshare.com/articles/Source\_data\_of\_Green\_s\_function\_and\_figure\_files\_/5131840/2}.

\clearpage

\end{document}